\documentclass[10pt]{article}

\usepackage{amsmath}
\usepackage{amssymb}

\def\b{\begin{equation}}
\def\e{\end{equation}}

\def\b{\begin{equation}}
\def\e{\end{equation}}
\def\bd{\begin{displaystyle}}
\def\ed{\end{displaystyle}}
\def\ba{\begin{array}}
\def\ea{\end{array}}

\def\bee{\begin{enumerate}}
\def\eee{\end{enumerate}}

\def\bes{\begin{eqnarray*}}
\def\ees{\end{eqnarray*}}
\def\be{\begin{eqnarray}}
\def\ee{\end{eqnarray}}

\begin{document}

\title{The Krein-Gupta-Bleuler Quantization in de Sitter Space-time;\\Casimir Energy-Momentum Tensor for a Curved Brane}
\author{S. Rahbardehghan$^1$ and H. Pejhan$^2$\thanks{e-mail: h.pejhan@piau.ac.ir}}
\maketitle
\centerline{\it $^1$Department of Physics, Islamic Azad University, Central Branch, Tehran, Iran}
\centerline{\it $^2$Department of Physics, Science and Research Branch, Islamic Azad University, Tehran, Iran}
\vspace{1.5pt}

\begin{abstract}
In this paper, vacuum expectation value (VEV) of the energy-momentum tensor for a conformally coupled scalar field in de Sitter space-time is investigated through the Krein-Gupta-Bleuler construction. This construction has already been successfully applied to the de Sitter minimally coupled massless scalar field and massless spin-2 field to obtain a causal and fully covariant quantum field on the de Sitter background. We also consider the effects of boundary conditions. In this respect, Casimir energy-momentum tensor induced by Dirichlet boundary condition on a curved brane is evaluated.
\end{abstract}

\section{Introduction}
The recent cosmological observations, which are strongly in favour of a positive acceleration of the present universe, can be well approximated by the de Sitter (dS) universe \cite{Riess}. Furthermore, de Sitter space-time plays an essential role in the inflationary scenario of the very early universe \cite{Hawking,Linde}. So, its metric becomes important at large-scale universe. The quantum field theory on dS space-time is also of considerable interest. De Sitter space is maximally symmetric curved space-time and offers the opportunity of controlling the transition to the flat space-time by the so-called contraction procedure, in which one can quantize fields and obtain simple exact solutions. However, even for this very simple space-time, this is not the case always. Allen has shown that, for the dS minimally coupled massless scalar field, which plays a central role in the inflationary models \cite{Linde2} and the linear quantum gravity in dS space, the covariant canonical quantization cannot be constructed over the Hilbert space because no invariant vacuum exists \cite{Allen}. Actually, the problem is originated due to the presence of a constant solution, the so-called zero mode problem; although this zero mode has positive norm, being part of the Hilbertian structure of the one particle sector is impossible. More accurately, regarding the conformal time, the action of the de Sitter group on this mode generates all the negative frequency solutions of the field equation. Therefore, the constructed Fock space over the Hilbert space (generated by any complete set of modes including the zero mode; ${\cal{H}}_{+}=\{\sum_{k\geq0} \alpha_k\phi_k; \sum_{k\geq0}|\alpha_k|^2<\infty \}$, $\phi_k$ is defined in \cite{Gazeau}) is not closed under the action of the dS group.

The failure of the usual canonical quantization in which the Fock space is constructed over the Hilbert space (the scalar product is positive) had also occurred in quantum electrodynamics. It is prevalently accepted that the origin of that impossibility is the invariance of the Lagrangian under a gauge transformation. It is proved that preserving covariance and gauge invariance in canonical quantization can be performed solely by exploiting the Gupte-Bleuler formalism.\footnote{In this formalism, $V_g$ stands for the space of gauge states (longitudinal photon states), while the space of positive frequency solutions of the field equation which satisfy the Lorentz condition is defined by $V$. Meanwhile, $V'$ is allotted to the all positive frequency solutions space which includes un-physical states. These spaces verify $V_g\subset V\subset V'$. The Fock space is constructed over $V'$, which is not a Hilbert space, but an indefinite inner product space; the Klein-Gordon inner product determines the Poincar\'{e} and locally and conformally invariant indefinite inner product on $V'$.  It should be noted that, all three spaces carry representations of the Poincar\'{e} group but $V_g$ and $V$ are not covariantly complemented. The quotient space $V/V_g$ of states up to a gauge transformation is the space of physical one-photon states (for more mathematical details, one can refer to \cite{BinegarGUPTA,GazeauGUPTA}).} Interestingly, the zero mode problem is deeply analogous to the QED case, for the Lagrangian
$${\cal{L}}=\sqrt{|g|}\partial_\mu \phi \partial^\mu \phi,$$
of the free massless field is invariant under $\phi\rightarrow\phi + \lambda$ ($\lambda$ is a constant function), which is similar to a gauge transformation. So, it would not be surprising if a canonical quantization of the Gupta-Bleuler type, would perform identically for dS massless minimally coupled scalar field; the set $\cal{N}$ of constant functions will serve as the space of gauge states, while $\cal{K}$ is a space of positive frequency solutions of the field equation equipped with the degenerate (but positive) Klein-Gordon inner product. However, the covariant quantum field cannot be constructed through a degenerate space of solutions \cite{Renaud}. Thus, by admitting $\cal{K}$ as an invariant subspace, a non degenerate invariant space of solutions $K$ must be built. These spaces, together with $\cal{N}$, are ingredients of the so-called Gupta-Bleuler triplet ${\cal{N}}\subset{\cal{K}}\subset K$. It is proved that $K$ is a Krein space, the direct sum of a Hilbert space and an anti-Hilbert space (a space with definite negative inner product). Indeed, the crucial point is originated in the fact that for the dS minimally coupled field a covariant decomposition, $ K = {\cal{H}}_{+} \oplus {\cal{H}}_{-}$, does not exist (though concerning the scalar massive field, such a decomposition exists, where ${\cal{H}}_{+}$ is the usual physical states space satisfying ${\cal{H}_{+}^*}={\cal{H}}_{-}$) \cite{Gazeau}. It is shown \cite{Gazeau,Renaud} that through the Krein-Gupta-Bleuler (KGB) structure, one can obtain a fully covariant construction of the minimally coupled quantum field on the de Sitter space-time. This field is interestingly free of infrared divergence.\footnote{Let us recall that the infrared problem on Minkowski space-time is due to the existence of solutions of arbitrary small frequencies. One could think that this problem, and the symmetry breaking associated with it, will disappear on the de Sitter space-time because the frequencies are now discretized since the space-time is spatially compact. But this is not quite true in the sense that, as we have just seen, the covariance of the theory forces one to include the null frequency solution itself in the normal mode decomposition of the field. This is of course in perfect agreement with Allen's result cited above.} The KGB method, therefore, provides a proposal to calculate graviton propagator on the dS background in the linear approximation, without any pathological behavior for largely separated points \cite{Takook1691,Behroozi,Dehghani,Garidi032501,Pejhan2052,TakookRouhani,Rahbardehghan}.

Motivated by these capabilities, in this paper another popular subject in this curved space-time, the interaction of fluctuating quantum fields with the background gravitational field and boundary conditions (Casimir effect), is investigated through the KGB structure. This effect demonstrates non-trivial charactristics of the quantum vacuum and has significant indications on all measures, from subnuclear to cosmological. In this regard and due to the importance of the braneworld scenarios in cosmology and particle physics ($\mbox{e.g.}$ see \cite{Wasserman,Randall}), we perform the calculation of the Casimir energy-momentum tensor in de Sitter space-time for a conformally coupled scalar field subjected to Dirichlet boundary condition on a curved brane.

The layout of the paper is as follows. In Sec. (2), we study the energy-momentum tensor for a conformally coupled scalar field in de Sitter space-time. Then in Sec. (3), the Casimir energy-momentum tensor in the presence of a curved brane as a boundary condition is computed. Finally we have enclosed the paper with a brief conclusion.

\section{Covariant Renormalization of the Energy-Momentum Tensor\\ through the KGB Structure}
To start, consider the $4+1$ dimensional (4+1-D) dS static coordinates, $x^i=(t,r,\theta,\vartheta,\phi)$,
\begin{equation} \label{dS} ds^2_{dS}=g_{ik}dx^i dx^k =(1-\frac{r^2}{\alpha^2})dt^2 - \frac{dr^2}{1-\frac{r^2}{\alpha^2}} - r^2d\Omega^2_3,\end{equation}
and a conformally coupled massless scalar field, $\phi(x)$, on this background that satisfies the equation
\begin{equation} \label{dS;Eq} (\nabla_l\nabla^l + \zeta R)\phi(x)=0, \;\;\zeta=\frac{3}{16}\end{equation}
in which $d\Omega^2_3$ is the line element on the $3$ dimensional unit sphere in Euclidean space, and the parameter $\alpha$ defines the dS curvature radius, $\nabla_l$  and $R$ are, respectively, the covariant derivative and the Ricci scalar for the corresponding metric. The inner product for the solutions space is defined as
\begin{equation}\label{inner}(\phi_1,\phi_2)=-i\int_\Sigma\phi_1\,{\mathop{\partial_\mu}\limits^\leftrightarrow}\,\phi^*_2d\Sigma^\mu,\end{equation}
where $d\Sigma^\mu=d\Sigma n^\mu$, and $d\Sigma$ is the volume element in a given space-like hypersurface, and $n^\mu$ is the time-like unit vector normal to this hypersurface. There exists a complete set of mode solutions of Eq. (\ref{dS;Eq}) which are orthonormal in the product (\ref{inner}), i.e.
\begin{equation} (\phi_k,\phi_{k'})=\delta _{kk'}, \;\;  (\phi_k^*,\phi_{k'}^*)= -\delta _{kk'}, \;\; (\phi_k,\phi_{k'}^*)= 0, \label{eq:ortho} \end{equation}
the set of $\{\phi_k ,\phi_k^*\}$ are, respectively, positive and negative norm states.

As already discussed, in order to have a fully covariant quantization scheme in dS space-time, utilizing the Krein-Gupta-Bleuler structure is unavoidable. In this construction, the field acts on a space of states having the structure of a Fock space but containing both positive and negative norm vectors. Actually, respecting the field equation (\ref{dS;Eq}) and its complete set of mode solutions (\ref{eq:ortho}), the field operator $\varphi$ would be as follows
\begin{equation}\label{123} \varphi =\frac{1}{\sqrt{2}} \Big( \sum_k (a_k \phi_k + a^\dagger_k \phi^*_k) + \sum_k (b^\dagger_k \phi_k + b_k \phi^*_k) \Big),\end{equation}
$a_k |0\rangle=0,\; b_k |0\rangle=0$ determine the Fock vacuum state $|0\rangle$, while $a_k^\dag |0\rangle=|1_k\rangle,\; \; b_k^\dag |0\rangle= |\bar 1_k\rangle$ are the physical and un-physical states. Note that, $[a_k, a^\dagger_{k'}] = \delta_{kk'},\; [b_k, b^\dagger_{k'}] = -\delta_{kk'}$ and the other commutation relations are zero.

With these definitions, it turns out that the field itself is not an observable: this is as expected and can be seen by calculating the mean value. The components of the energy-momentum tensor on the other hand are observables. To see the point, consider the operator $T_{\mu\nu}$ which obviously is not positively definite as an operator on the full space of states. In order to compute expectation value of the energy-momentum tensor, $\langle \vec{k}|T_{\mu\nu}| \vec{k}\rangle$, in which $|\vec{k}\rangle$ is the excited physical state
\begin{equation} |\vec{k}\rangle \equiv |{k}_1^{n_1}...{k}_j^{n_j}\rangle = \frac{1}{\sqrt{n_1!...n_j!}}(a_{k_1}^\dag)^{n_1}...(a_{k_j}^\dag)^{n_j}|0\rangle, \end{equation}
one should generally begin with
\begin{equation} \langle \vec{k}|\partial_\mu\varphi(x)\partial_\nu\varphi(x)|\vec{k}\rangle = \sum_k \partial_\mu\phi_k(x)\partial_\nu\phi_k^\ast(x) - \partial_\mu\phi_k^\ast(x)\partial_\nu\phi_k(x) + 2 \sum_i n_i \Re \Big(\partial_\mu\phi_{k_i}^\ast(x)\partial_\nu\phi_{k_i}(x)\Big). \end{equation}
In analogy with the conventional approach, the first term is responsible for the appearance of infinite divergences in the theory. However, in the KGB approach, the unusual presence of the second term with the minus sign which comes from the terms of the field including $b_k$ and $b_k^\dag$, can automatically remove this term. Therefore, we have
\begin{equation} \langle \vec{k}|\partial_\mu\varphi(x)\partial_\mu\varphi(x)|\vec{k}\rangle = 2 \sum_i n_i \partial_\mu\phi_{k_i}^\ast(x)\partial_\mu\phi_{k_i}(x). \end{equation}
Correspondingly, there exists an automatic renormalization of the $T_{\mu\nu}$'s (no infinite term appears). A straightforward result of this construction, which assures a reasonable physical interpretation of the model, is the positivity of the energy for any physical state $|\vec{k}\rangle$; $\langle \vec{k}|T_{00}| \vec{k}\rangle\geq0$ ($ = 0 \Leftrightarrow |\vec{k}\rangle=|0\rangle$).

In addition, this procedure fulfills the so-called Wald axioms:
\begin{itemize}
\item{The causality and covariance is assured since the field is.}
\item{For physical states, with respect to the above calculation, it turns out that the formal results are preserved.}
\item{The foundation of the above computation is as follows (note that $[b_k,b_k^\dag]=-1$)
$$a_ka_k^\dag+a_k^\dag a_k+b_kb_k^\dag+b_k^\dag b_k=2a_k^\dag a_k+2b_k^\dag b_k,$$
Which provides a reordering equivalent when the method is applied to physical states (on which $b_k$ vanishes).}
\end{itemize}

The method, therefore, presents an interesting property linked to the vacuum energy issue in curved space-time. To have a deeper insight into the subject from the viewpoint of the KGB approach, let us reconsider the field equation (\ref{123}) in a more explicit form as follows
\begin{equation} \varphi = \sum_k \left(A_k \phi_k + A_k^\dag \phi^*_k\right), \;\;\; \mbox{in which}\; A_k\equiv\frac{a_k+b^\dagger_k}{\sqrt{2}}. \label{fieldcurved1} \end{equation}
Note that, the operators $A_k$ no longer verify $A_k|0\rangle=0$. Nevertheless, by using the operator ${\cal{D}}_k = A_k A_k^\dag + A_k^\dag A_k$, one can determine the Fock vacuum state as (the point is $[b_k,b_k^\dag]=(\phi_k^\ast,\phi_k^\ast)=-1$)
\begin{equation}\langle 0|{\cal{D}}_k|0\rangle =0, \;\;\;\forall k.\label{counter} \end{equation}
Interestingly, it is proved that this equation is independent of Bogolubov transformations \cite{HawkingRadiation}. So in this method, in contrast to the usual approach where the vacuum is determined through the modes and it is usually said that the choice of the modes is equivalent to the choice of the vacuum, the Fock vacuum is unique and therefore does not specify the physical space of states. However, this does not mean that the Bogolubov transformations which only modify the set of physical states are no longer valid in this construction. Any physical state depends on the selected space-time and also on the observer; the physical states of an accelerated observer in Minkowski space are different from those of an inertial observer (Unruh effect) \cite{Garidi}. While, the same representation of the field can be employed for both cases (it is invariant under Bogolubov transformations). Indeed, instead of having a multiplicity of vacua, we have several possibilities for the space of physical states, so the usual ambiguity about vacua is not suppressed but displaced.

Note that, due to the automatic renormalization of the $\langle T_{\mu\nu} \rangle$ through this construction, the expected value of all components of the energy-momentum tensor vanish in the vacuum, and hence there is no conformal anomaly in the trace of the energy-momentum tensor. From this point of view, this renormalization scheme seems to be very different from the other ones which all present this anomaly. Of course, it is not very surprising that our field, which is covariant and conformally covariant in a rather strong sense \cite{Renaud}, does not present any conformal anomaly which, after all, can appear only by breaking the conformal invariance.

In the next section, with respect to the importance of braneworld scenarios, we calculate the Casimir energy-momentum tensor for a curved brane through the KGB method.

\setcounter{equation}{0}
\section{Considering the Theory in the Presence of Boundary Condition}
In this section, by considering the de Sitter space-time as the gravitational background, we evaluate the Casimir energy-momentum tensor for the conformally coupled scalar field, see (\ref{dS;Eq}) and (\ref{123}), subjected to Dirichlet boundary condition on the hypersurface $S$ (we will determine the explicit form of the hypersurface in the next few lines). Technically, to make the maximum use of the flat space-time calculations, we present the dS line element (\ref{dS}) in the form conformally related to the Rindler metric. Regarding the coordinate transformation; $ x^i \rightarrow {x'^i}= (\eta,\sigma,X'),\; X'=({x'^2},x'^3,x'^4)$,
\begin{equation}\label{1} \eta=\frac{t}{\alpha}, \;\; \sigma= \frac{\sqrt{\alpha^2-r^2}}{\Omega},\;\; (\mbox{in which}\;\Omega=1-\frac{r}{\alpha}\cos\theta) \end{equation}
\begin{equation}\label{2} x'^2=\frac{r}{\Omega}\sin\theta\cos\vartheta,\;\; x'^3=\frac{r}{\Omega}\sin\theta\sin\vartheta\cos\phi,\;\;x'^4=\frac{r}{\Omega}\sin\theta\sin\vartheta\sin\phi. \end{equation}
the dS line element (\ref{dS}) takes the form
\begin{equation}\label{3} ds^2_{dS}=g'_{ik}dx'^idx'^k = \Omega^2 (\sigma^2d\eta^2- d\sigma^2- dX'^2 ), \end{equation}
that is manifestly conformally related to the Rindler space-time
\begin{equation}\label{4} ds^2_{dS}= \Omega^2ds^2_{R},\;\; ds^2_{R}= \bar{g}_{ik}dx'^idx'^k= \sigma^2d\eta^2- d\sigma^2- dX'^2,\;\; g'_{ik}= \Omega^2 \bar{g}_{ik}. \end{equation}

As the boundary condition, we take an infinite plane moving by uniform acceleration normal to itself which can be determined by the coordinate $\sigma=b$ in the right Rindler wedge. Note that, the curves $\sigma=\mbox{constant}$, $X'=\mbox{constant}$ are worldlines of constant proper acceleration $\sigma^{-1}$ and the surface $\sigma=b$ represents the trajectory of the barrier which has a proper acceleration $b^{-1}$. In the dS static coordinates the boundary $S$ is presented by the following curved brane
\begin{equation}\label{5} \sqrt{\alpha^2 - r^2}= b(1-\frac{r}{\alpha}\cos\theta). \end{equation}

As a Rindler counterpart one can consider the vacuum energy-momentum tensor induced by $S$ (as Dirichlet boundary condition is conformally invariant, the Dirichlet scalar in the curved bulk corresponds to the Dirichlet scalar in a flat space-time). Accordingly, the problem under consideration would be a conformally trivial situation; a conformally invariant field on background of the conformally flat space-time. So, instead of evaluating Casimir energy-momentum tensor directly on dS background, with regard to the standard transformation formula for the VEV of the energy-momentum tensor in conformally related problems \cite{Birrell}, one can generate the results for dS space-time from the corresponding results for the Rindler space-time.

In this regard and in the beginning, we should pursuit quantizing procedure in Minkowski space-time for the massless scalar field, $\Box\phi(x)=0$,
for which the inner product of a pair of its solutions is defined by
\begin{equation}(\phi_1,\phi_2)=-i\int (\phi_1(x) \, {\mathop{\partial_\mu}\limits^\leftrightarrow } \,\phi^*_2(x)) d^3 x.\end{equation}
As already discussed, the field operator in the KGB quantization would be $\varphi = \frac{1}{\sqrt{2}}({\varphi}_+ + {\varphi}_-)$, in which
\begin{equation}\begin{aligned}\label{fk}
\varphi_+(x)=\int d^3 \vec{k}\;[a(\vec{k})\phi(\vec{k},x)+a^\dag(\vec{k})\phi^\ast(\vec{k},x)],\\
\varphi_-(x)=\int d^3 \vec{k}\;[b(\vec{k})\phi^\ast(\vec{k},x)+b^\dag(\vec{k})\phi(\vec{k},x)],
\end{aligned}\end{equation}
where $\varphi_+(x)$ and $\varphi_-(x)$ are, respectively, physical and un-physical part of the field operator. $\phi (\vec{k},x)=({4\pi\omega})^{-1/2}{e^{i\vec{k}\cdot\vec{x}-i\omega t}}$ and $[a(\vec{k}), a^\dagger(\vec{k}')] = \delta (\vec{k}-\vec{k}'),\;\; [b(\vec{k}), b^\dagger(\vec{k}')] = -\delta (\vec{k}-\vec{k}')$, the other commutation relations are zero. The Fock vacuum state $|0\rangle$ is defined by $a(\vec{k}) |0\rangle=0,\; b(\vec{k}) |0\rangle=0$.

Due to the presence of un-physical states in the KGB structure, when interacting fields are investigated, the unitarity of the S-matrix must be preserved. This would be obtained by the following procedure, which is the so-called unitarity condition \cite{Garidi}; let $\Pi_+$ be the projection over ${\cal{H}}_+$
$$\Pi_+ = \sum_{\{\alpha_+\}} |\alpha_+><\alpha_+|,\;\;\;\;\; |\alpha_+>\;\in {\cal{H}}_+.$$
So, considering the field operator, one has
\begin{equation}\label{unicon} \Pi_+ \varphi \Pi_+ |\alpha > = \left\{\begin{array}{rl} &\varphi_+ |\alpha >, \;\;\;\;\;\; \mbox{if}\; |\alpha>\;\in {\cal{H}}_+ \vspace{2mm}\\\vspace{2mm} &0,\;\;\;\;\;\;\;\;\;\;\;\;\;\;\;\;\mbox{if}\; |\alpha>\;\in {\cal{H}}_{-} \\\end{array}\right. \end{equation}
Correspondingly, in place of a standard selection for the Lagrangian potential term, $V(\varphi)$, we consider the restricted form of $V$ to the positive energy modes as $V'(\varphi)\equiv V(\Pi_+\varphi\Pi_+)$. As a result, vacuum influences in the theory only include the interacting vacuum.

Therefore, with regard to the unitarity condition, the influence of applying physical boundary condition on the field operator is only upon the physical states. Accordingly, when physical boundary conditions are present, the field operator would be as
\begin{equation}\label{effected} \varphi(x)=\sum_d [a({\vec{k}}_d) \phi(\vec{k}_d,x)+a^\dag({\vec{k}}_d)\phi^\ast(\vec{k}_d,x)] + \int d^3 \vec{k}\; [b(\vec{k})\phi^\ast (\vec{k},x)+b^\dag (\vec{k})\phi (\vec{k},x)], \end{equation}
here $\vec{k}_d$ are the eigen-frequencies of the system under consideration.

Before calculating the energy-momentum tensor in view of the accelerated (Rindler) coordinates (defined by the coordinate transformation: $t=\sigma\sinh\eta$, $x=\sigma\cosh\eta$, which cover the region $|x|>|t|$ of Minkowski space), the construction of the Feynman Green function of the KGB method, $G(x,x')$, in the Rindler space is required. In this regard, it is easily seen that with respect to the quantum field of the theory, (\ref{effected}), it can be decomposed into two parts, physical and un-physical parts (the point is $(\phi_k,\phi_{k'}^*)= 0$ and $[a(k), b^\dagger(k')] = [a(k),b(k')] = 0$). Accordingly, we have
\begin{equation}\label{G} G(x,x')= G_{+}(x,x') + G_{-}(x,x'). \end{equation}
Here, the physical part of the theory which is subjected to Dirichlet boundary condition is determined by $G_{+}(x,x')$, while $G_{-}(x,x')$ refers to the un-physical part of the theory which according to its definition would be $G_{-}(x,x')= - G_{0}(x,x')$, where $G_{0}(x,x')$ is the Feynman propagator for a free massless scalar field on the entire Minkowski manifold and the sign "$-$" is due to $[b_k,b_k^\dag]=(\phi_k^\ast,\phi_k^\ast)=-1$. The corresponding propagators for the Dirichlet and the free field ones have been computed, respectively, in Refs. \cite{Candelas} and \cite{Candelas'}. So, for our considered case we have
\begin{equation}\label{DP} G_{+}(x,x')= G_{0}(x,x') - \frac{i}{\pi}\int \frac{d\nu}{2\pi} \exp [-i\nu(\eta-\eta')]\int \frac{d^2k}{(2\pi)^2}\exp [ik\cdot (x-x')] \frac{K_{i\nu}(e^{i\pi}kb)}{K_{i\nu}(kb)} K_{i\nu}(k\sigma) K_{i\nu}(k\sigma'), \end{equation}
\begin{equation}\label{DP'}G_{-}(x,x') \Big(= - G_{0}(x,x')\Big) = -\frac{i}{\pi}\int \frac{d\nu}{2\pi} \exp [-i\nu(\eta-\eta')] \int \frac{d^2k}{(2\pi)^2}\exp [ik\cdot (x-x')] K_{i\nu}(k\sigma_>) K_{i\nu}(e^{i\pi}k\sigma_<), \end{equation}
in which $K_{i\nu}(k\sigma)$ is the modified Bessel function of imaginary order.

Now respecting the Feynman Green function (\ref{G}), the VEV of the energy-momentum tensor in view of the accelerated observer in Minkowski space-time would be
\begin{equation} \label{st} <0|T_\mu^\nu|0>^{Rindler}=-i\lim_{x'\rightarrow x} (\frac{2}{3}\nabla_{\mu}\nabla^{\nu'}- \frac{1}{3}{\nabla_{\mu}}\nabla^{\nu}- \frac{1}{6}g_\mu^\nu\nabla_{\alpha}\nabla^{\alpha'})G(x,x').\end{equation}
Considering (\ref{DP}) and (\ref{DP'}) and after a straightforward calculations, we have
\begin{equation}\label{diagT} <T_\mu^\nu>^{Rindler}=\mbox{diag}(A,B,C,C), \end{equation}
where
\begin{equation}\label{abg}
\left\{\begin{array}{rl} \begin{aligned}
A(\sigma)&=\frac{1}{12\pi^3}\int_0^{\infty} d\nu\int_0^{\infty} dk\;k\frac{K_{i\nu}(e^{i\pi}kb)}{K_{i\nu}(kb)}\left[ k^2K_{i\nu}^{'2}(k\sigma)+\frac{2k}{\sigma}K_{i\nu}(k\sigma)K'_{i\nu}(k\sigma)+\left( \frac{5\nu^2}{\sigma^2}+k^2 \right)K_{i\nu}^2(k\sigma) \right],\\
B(\sigma)&=-\frac{1}{12\pi^3}\int_0^{\infty} d\nu\int_0^{\infty} dk\;k\frac{K_{i\nu}(e^{i\pi}kb)}{K_{i\nu}(kb)}\left[ 3k^2K_{i\nu}^{'2}(k\sigma)+\frac{2k}{\sigma}K_{i\nu}(k\sigma)K'_{i\nu}(k\sigma)+3\left( \frac{\nu^2}{\sigma^2}-k^2 \right)K_{i\nu}^2(k\sigma) \right],\\
C(\sigma)&=-\frac{1}{12\pi^3}\int_0^{\infty} d\nu\int_0^{\infty} dk\;k\frac{K_{i\nu}(e^{i\pi}kb)}{K_{i\nu}(kb)}\left[ -k^2K_{i\nu}^{'2}(k\sigma)+\left( \frac{\nu^2}{\sigma^2}+2k^2 \right)K_{i\nu}^2(k\sigma) \right].
\end{aligned}
\end{array}\right.
\end{equation}
It is trace-free, $A+B+2C=0$, and conserved, $A=\frac{d}{d\sigma}(\sigma B)$. We should emphasize that, in Ref. \cite{Candelas} by P. Candelas \textit{et al}., the above regularized result was obtained by subtracting the value that it would have if evaluated relative to the Minkowski vacuum. In our method, however, the VEV of the energy-momentum tensor is automatically regularized.

From now on, therefore, in analogy with the procedure that was proposed in \cite{Candelas} to evaluate the above integrals, one can find:
\begin{itemize}
\item {The asymptotic form, $\sigma/b \rightarrow\infty$, for $T_{\mu}^{\nu}$ in the frame of an observer with proper acceleration $\sigma^{-1}$ as follows
\begin{equation}\label{MM} <T_\mu^\nu>^{Rindler}\sim \frac{-1}{2\pi^2\sigma^4} \int_0^\infty \frac{\nu^3 d\nu}{e^{2\pi\nu}-1}diag(-1,\frac{1}{3},\frac{1}{3},\frac{1}{3}), \end{equation}
Which present a negative energy density with a  Planckian spectrum with the temperature $T=(2\pi\sigma)^{-1}$, corresponding to the absence from the vacuum of black-body radiation.  This result is independent of the particular boundary conditions and of the acceleration of the barrier in the sense that it depends only on the acceleration $\sigma^{-1}$ of the local Killing trajectory. It means that regarding the gravitational analogy, at sufficiently large distances to the barrier, VEV of the energy-momentum tensor depends purely on the local gravitational field.

Note that, when $\sigma$ is analogous to $b$, one would expect the deviation from a black-body spectrum since the temperature corresponds to a wavelength comparable with the distance from the barrier. So, the effects of the boundary conditions become important; the rate at which these effects decline far from the barrier can be found by proceeding to the next order in the asymptotic expansion of $T_\mu^\nu$. It turns out that the terms dependent on the boundary condition are of order $(\sigma^4\mbox{ln}^3\sigma)^{-1}$
\begin{equation}\label{appT} T_\mu^\nu=[ -(480\pi^2 \sigma^4)^{-1}-(144\sigma^4\mbox{ln}^3\sigma/b)^{-1}]\mbox{diag}(-1,\frac{1}{3},\frac{1}{3},\frac{1}{3})+{\cal {O}}[(\sigma^4 \mbox{ln}^4\sigma/b)^{-1}]\;\;\;\;\;(\sigma/b\longrightarrow\infty)\end{equation}}

\item{At the moderate distance, further analytical simplification cannot be applied to the integral representations (\ref{abg}). Therefore, the numerical quadrature should be used to evaluate $T_\mu^\nu$ in this intermediate region. It is found that the energy density is always negative and decreases monotonically towards the barrier \cite{Candelas}.}

\item{In the limit of small separation from the barrier, $\sigma/b\longrightarrow1$, boundary conditions are not effective and it is again possible to compute the explicit asymptotic forms for the energy-momentum tensor which has been shown to vary as the inverse cube of the distance from the barrier, and is therefore unbounded.
    \begin{equation}\label{appabg}
\left\{\begin{array}{rl} \begin{aligned}
A &=[360\pi^2b(\sigma-b)^3]^{-1}+{\cal {O}}[(\sigma/b-1)^{-2}],\\
B &=-[720\pi^2b^2(\sigma-b)^2]^{-1}+{\cal {O}}[(\sigma/b-1)^{-1}],\\
C &=-[720\pi^2b(\sigma-b)^3]^{-1}+{\cal {O}}[(\sigma/b-1)^{-2}].
\end{aligned}
\end{array}\right.
\end{equation}}
\end{itemize}
Note that, the above results are relevant to the right Rindler wedge. By imposing the replacements $I_{i\nu}\longrightarrow K_{i\nu}, K_{i\nu}\longrightarrow I_{i\nu}$ in Eq. (\ref{abg}), one can straightforwardly obtain the expression for the boundary part of the vacuum energy-momentum tensor in the region $\sigma<b$.

The vacuum energy-momentum tensor on the static de Sitter space Eq. (\ref{dS}), therefore, can be easily obtained by the standard transformation law between conformally related problems as follows
\begin{equation}\label{dST} <T_\mu^\nu>^{dS}= \Omega^{-5} <T_\mu^\nu>^{Rindler}. \end{equation}
Now considering Eq. (\ref{G}), one can easily realize that $G(x,x')$ is finite for $\sigma>b$ since the singular parts of $G_{+}$ and $G_{-}$ cancel as $x'\longrightarrow x$. However, if $x$ is a point on this brane then $$G(x,x')= G_{+}(x,x') + G_{-}(x,x')=-G_{0}(x,x'),$$ for $G_+$ vanishes at the barrier due to the choice of Dirichlet boundary condition. If we now let $x'\longrightarrow x$, we find that $G(x,x')$ exhibits the same singularity as the Minkowski space propagator $i\left[ 4\pi^2(x-x')^2 \right]^{-1}$. Regarding this reasoning $T_\mu^\nu$ could, \textit{a priori}, diverge as $(\sigma -b)^{-4}$. However, when the acceleration of the barrier is reduced to zero, $b\longrightarrow \infty$, the energy-momentum tensor must vanish. So, the reproduction of the result for a single barrier at rest entails the leading behavior of $T_\mu^\nu$, as is found, be at worst of order $\left[ b(\sigma-b)^3 \right]^{-1}$ \cite{Candelas}. Note that, in order to have a finite $T_\mu^\nu$ there, accurate cancellations are required, which do not occur for massless scalar field subjected to Dirichlet boundary condition. Nevertheless, selecting such a field, apart from the braneworld models motivation, is due to the relevant important considerations of this field which greatly encourage one to pursuit this path to study the electromagnetic field.

\textit{Remarks on the renormalization:} It is worth to mention that in the standard QFT (the Fock space is built on a Hilbert space), any prescription for renormalizing the energy-momentum tensor, which is consistent with Wald axioms, must yield precisely the trace, modulo the trace of a conserved local curvature term \cite{Wald}. And for the studied case in this paper, a conformally coupled quantum field on the conformally flat background, as has been proved in \cite{Birrell}, the gravitational part of the energy-momentum tensor is entirely determined by the trace anomaly. The above calculations, however, reveal that the manifestation of the gravitational background (the trace anomaly), which corresponds to the situation without boundaries, does not appear. It is not surprising since the KGB structure preserves covariant and conformally covariant in a rather strong sense, therefore the trace anomaly, which can appear only through the conformal anomaly (breaking the conformal invariance when quantum corrections are included), vanishes. In this respect, we should also underline that although the trace anomaly is absent, Wald axioms are well preserved in the context of the KGB construction. Moreover, the covariant automatic renormalization of the energy-momentum tensor, as a significant achievement of this theory, can be helpful in further studying of quantum field theory in curved space-time, where the usual scheme of renormalization includes complexity and somewhat ambiguity.

\section{Discussion}
In this paper, we discussed the bulk Casimir effect for a conformal scalar field when the bulk represents 5-D de Sitter space-time with one 4-D dS brane, which may be similar to our universe. Due to the fact that the braneworld corresponds to a 5-D manifold with a 4-D dynamical boundary, it is obvious that, regarding the 5-D quantum field theory, the non-trivial vacuum energy should appear. Furthermore, in the context of the brane QFT, the non-trivial brane vacuum energy also emerges. In this regard, the bulk Casimir effect should conceivably serve as the cornerstone in the construction of the consistent braneworlds. Indeed, it gives contribution to both the brane and the bulk cosmological constants. Therefore, one expects that it can be helpful in the resolution of the cosmological constant problem. Indeed, as almost commonly accepted, this problem is mainly a question about quantum gravity, since classically it would be more or less natural to just decide - as Einstein did - that we do not like the cosmological constant, and set it to zero. Accordingly, in the investigation and resolution of the cosmological constant problem the inclusion of the dynamics of quantum gravity would be a crucial step \cite{Govaerts,Witten}.

Accordingly, the Krein-Gupta-Bleuler structure is considered in this paper to perform the computations. The authors would like to emphasize the fact that the method not only fulfills the so-called Wald axioms but also is consistent with the de Sitter linear gravity requirements; the causality and the covariance of the theory are assured and contrary to what happened in previous treatments of this problem, the model does not suffer from infrared divergences \cite{Takook1691,Behroozi,Dehghani,Garidi032501,Pejhan2052,TakookRouhani,Rahbardehghan}. Here, it is also worth to mention that, as accurately discussed in Ref. \cite{Garidi}, considering the unitarity condition (see (\ref{unicon}) and related discussion) when interaction is present - in the Minkowskian limit - the behaviour of the method, more precisely, the so-called radiative corrections are the same as the usual QFT. Indeed, in the flat limit, vanishing of the free field vacuum energy in the KGB structure is the only difference between our approach and the usual one. In this regard, we also showed that utilizing the method does not destroy black holes thermodynamics and it is able to retrieve the very result for Hawking radiation even regarding the fact that $<0|T_\mu^\nu|0>$ of the free theory is zero \cite{HawkingRadiation}. In this respect, we hope that the current study based on the Krein-Gupta-Bleuler construction can be extended to the related issues, such as calculation of the Casimir force, due to quantum gravity in dS space.


\begin{thebibliography}{a}
\addcontentsline{toc}{chapter}{Bibliographie}

\bibitem{Riess} A.G. Riess et al. [Supernova Search Team Collaboration], Astro. J 116, 1009 (1998); S. Perlmutter et al. [Supernova Cosmology Project Collaboration], Astro. J 517, 567 (1999); U. Seljak, A. Slosar, and P. McDonald, JCAP 014, 610 (2006); A.G. Riess et al., Astro. J 98, 659 (2007).

\bibitem{Hawking} \emph{The Very Early Universe} edited by G.W. Gibbons, S.W. Hawking, and S.T.C. Siklos (Cambridge University Press, London, 1983).

\bibitem{Linde} A.D. Linde, \emph{Particle Physics and Inflationary Cosmology}, Harwood Academic, Chur, (1990).

\bibitem{Linde2} A. Linde, Phys. Lett. B 116, 335 (1982).

\bibitem{Allen} B. Allen, Phys. Rev. D 32, 3136 (1985); B. Allen, A. Folacci, Phys. Rev. D 35, 3771 (1987).

\bibitem{Gazeau} J.P. Gazeau, J. Renaud, and M.V. Takook, Class Quant. Grav. 17, 1415 (2000).

\bibitem{BinegarGUPTA} B. Binegar, C. Fronsdal and W. Heidenreich, J. Math. Phys. 24, 2828 (1983).

\bibitem{GazeauGUPTA} J.P. Gazeau, J. Math. Phys. 26, 1847 (1985).

\bibitem{Renaud} S. De Bi\'{e}vre and J. Renaud, Phys. Rev. D 57, 10 (1998).

\bibitem{Takook1691} M.V. Takook, Mod. Phys. Lett. A. 16, 1691 (2001).

\bibitem{Behroozi} S. Behroozi, S. Rouhani, M.V. Takook and M.R. Tanhayi, Phys. Rev. D 74, 124014 (2006).

\bibitem{Dehghani} M. Dehghani, S. Rouhani, M.V. Takook and M.R. Tanhayi, Phys. Rev. D 77, 064028 (2008).

\bibitem{Garidi032501} T. Garidi, J.P. Gazeau, S. Rouhani and M.V. Takook, J. Math. Phys 49, 032501 (2008).

\bibitem{Pejhan2052} M.V. Takook, H. Pejhan, M.R. Tanhayi, Eur. Phys. Jour. C 72, 2052 (2012).

\bibitem{TakookRouhani} M.V. Takook and S. Rouhani, \emph{Quantum Linear Gravity in de Sitter Universe I: On Gupta-Bleuler vacuum state}, arXiv:1208.5562.

\bibitem{Rahbardehghan} S. Rahbardehghan, H. Pejhan, M. Elmizadeh, Eur. Phys. Jour. C 75, 119 (2015).

\bibitem{Wasserman} S.-H. Henry Tye, Ira Wasserman, Phys. Rev. Lett. 86, 1682 (2001).

\bibitem{Randall} L. Randall and R. Sundrum, Phys. Rev. Lett. 83, 3370 (1999).

\bibitem{HawkingRadiation} H. Pejhan and S. Rahbardehghan, arXiv:1408.4531 \emph{Krein Quantization Approach to Hawking Radiation}.

\bibitem{Garidi}  T. Garidi, E. Huguet and J. Renaud, J. Phys. A 38, 245 (2005).

\bibitem{Birrell} N.D. Birrell, P.C.W. Davies, Cambridge University Press, (1982), \emph{Quantum Field in Curved Space}.

\bibitem{Candelas} P. Candelas and D. Deutsch, Proc. Roy. Soc. A 79, 354 (1977).

\bibitem{Candelas'} P. Candelas and D.J. Raine, J. Math. Phys 17, 2101 (1976).

\bibitem{Wald} R.M.Wald, Phys. Rev. D 17, 1477 (1978).

\bibitem{Govaerts} J. Govaerts and S. Zonetti, Phys. Rev. D 87, 084016 (2013).

\bibitem{Witten} E. Witten, arXiv:hep-ph/0002297 \emph{The Cosmological Constant From The Viewpoint Of String Theory}.


\end{thebibliography}
\end{document}